\documentclass[prb,twocolumn,showpacs,superscriptaddress]{revtex4}

\usepackage{graphicx}
\usepackage{dcolumn}
\usepackage{amsmath2000}
\usepackage{amssymb}
\usepackage{bm}

\newcommand{\bea}{\begin{eqnarray}}
\newcommand{\eea}{\end{eqnarray}}
\newcommand{\ba}{\begin{array}}
\newcommand{\ea}{\end{array}}

\newcommand{\charge}{q}
\newcommand{\ladeta}{a_{\charge}}
\newcommand{\Eqref}[1]{Eq.~\ref{#1}}
\newcommand{\Figref}[1]{Fig.~\ref{#1}}
\newcommand{\Citeref}[1]{Ref.~\onlinecite{#1}}
\newcommand{\etal}{\textit{et~al.}} 
\newcommand{\bvec}[1]{\mathrm{\mathbf{#1}}}
\newcommand{\typeI}{type-I~}
\newcommand{\typeII}{type-II~}
\newcommand{\typeIII}{type-I/II~}
\newcommand{\xct}{x_{\text{tri}}}
\newcommand{\kct}{\kappa_{\text{tri}}}
\newcommand{\yct}{y_{\text{tri}}}
\newcommand{\ns}  {$\circ$}    
\newcommand{\sok} {$\star$}    
\newcommand{\ssim}{$\bullet$}  
\newcommand{\RR}{\overline{\left|\psi\right|^2}}
\newcommand{\deltaR}{\Delta \RR}

\begin{document}

\title{The order of the metal to superconductor transition} 

\author{S. Mo} 
  \email{Sjur.Mo@phys.ntnu.no}
  \affiliation{%
    Department of Physics \\
    Norwegian University of Science and Technology, \\
    N-7491 Trondheim, Norway} 
\author{J. Hove}
  \email{Joakim.Hove@phys.ntnu.no}
  \affiliation{%
     Department of Physics \\
     Norwegian University of Science and Technology, \\
     N-7491 Trondheim, Norway}
\author{A. Sudb{\o}}
  \email{Asle.Sudbo@phys.ntnu.no}
  \affiliation{%
     Department of Physics \\
     Norwegian University of Science and Technology, \\
     N-7491 Trondheim, Norway}
  \affiliation{%
     Institut f{\"u}r Theoretische Physik \\
     Freie Universit{\"a}t Berlin \\
     Arnimallee 15, D-14195 Berlin, Germany} 

\date{\today}

\pacs{74.55.+h,74.60.-w, 74.20.De, 74.25.Dw}

%************************************************************************************

\begin{abstract}
  We present results from large-scale Monte Carlo simulations on the
  full Ginzburg-Landau (GL) model, including fluctuations in the
  amplitude and the phase of the matter-field, as well as fluctuations
  of the non-compact gauge-field of the theory. {}From this we obtain
  a precise critical value of the GL parameter $\kct$ separating a
  first order metal to superconductor transition from a second order
  one, $\kct = (0.76\pm 0.04)/\sqrt{2}$. This agrees surprisingly well
  with earlier analytical results based on a disorder theory of the
  superconductor to metal transition, where the value
  $\kct=0.798/\sqrt{2}$ was obtained. To achieve this, we have done
  careful infinite volume and continuum limit extrapolations. In
  addition we offer a novel interpretation of $\kct$, namely that it
  is also the value separating \typeI and \typeII behaviour.
\end{abstract}

\maketitle

%***********************************************************************************

\section{Introduction}

The character of the metal to superconductor transition is an
important and long-standing problem in condensed matter physics.  The
critical properties of a superconductor may be investigated at the
phenomenological level by the Ginzburg-Landau (GL) model of a complex
scalar matter field $\phi$ coupled to a fluctuating mass-less
gauge-field $\bvec{A}$. The GL model in $d$-dimensions is defined by
the functional integral
\begin{align}
 Z &= \int \mathcal{D}A_i \mathcal{D}\phi \exp(-S(A_i,\phi)) \nonumber \\
 S &= \int d^dx  \left[\frac{1}{4}  F_{ij}^2 + |D_i \phi|^2 
+ m^2 |\phi|^2 + \lambda |\phi|^4 \right] 
\label{GLM}
\end{align}   
where $F_{ij} = \partial_i A_j - \partial_j A_i$, $D_i = \partial_i +
i \charge A_i$, $\charge$ is the charge coupling the condensate matter
field to the fluctuating gauge-field, $\lambda$ is a self-coupling,
and $m^2$ is a mass parameter which changes sign at the mean field
critical temperature. This model is also used to describe a great
number of other phenomena in Nature, including such widely separated
phenomena as the Higgs mechanism in particle physics,~\cite{Coleman:1973}
phase transitions in liquid crystals,~\cite{Gennes:1972,Lubensky:1978} crystal
melting,~\cite{Kleinert:1989book2} the quantum Hall
effect,~\cite{Wen:1993,Pryadko:1994} and it is also used as an effective field
theory describing phase transitions in the early Universe.~\cite{Vilenkin:1994book}

The GL model may conveniently be formulated in terms of two
dimensionless parameters $y= m^2/\charge^4$ and
$x=\lambda/\charge^2$ when all dimensionful quantities are expressed
in powers of the scale $\charge^2$. Here, $y$ is temperature-like and
drives the system through a phase transition, and $x = \kappa^2$ is
the well known GL parameter.  These parameters are related to the
standard dimensionful textbook~\cite{Tinkham:1996book} coefficients
$\alpha,\beta$ of the GL model by
\begin{equation}
\label{eq:coeff}
        y = \frac{m^{\ast}c^2}{128\pi^2 \alpha_s^2 k_{\text{B}}^2T^2} \alpha,\quad
        x = \frac{1}{8\pi\alpha_s \hbar c}\left(\frac{m^{\ast}c}{\hbar}\right)^2 \beta
          = \kappa^2
\end{equation}
where $\alpha_s$ is the fine structure constant%
\footnote{%
The fine structure constant is given by
$\alpha_s=\frac{\mu_0 c {e^{\ast}}^2}{4\pi\hbar}$
where $2e^{\ast}$ is the \emph{effective} charge of a Cooper pair.} 
and $m^{\ast}$ is an effective mass parameter. 

At the mean-field level \Eqref{GLM} reduces to the well known
GL-equations and the model exhibits a second order phase transition when
the temperature (or $y$) is varied through some critical value.  In a seminal 
paper by Halperin,
Lubensky and Ma~\cite{Halperin:1974} it was shown that by ignoring
spatial fluctuations in $\phi$, and then integrating out the $\bvec{A}$
field \emph{exactly}, one gets a term $\left| \phi \right|^3$ in the
effective $\phi$ action. Treating this action at the mean field level
leads to the prediction of a first order transition in the charged 
model for any value of the charge, or equivalently for any value of 
the GL parameter.
The first order character of the transition is most strongly pronounced
for large values of the charge (small $\kappa$), but even then it is very 
weak.  For $\kappa \ll 1$ (type-I) the neglect of spatial variation in the
matter field $\phi$ is a
reasonable approximation, whereas for $\kappa \gtrsim 1$ (type-II)
fluctuations in $\phi$ must be taken into account. By doing a one-loop
RG calculation using $\varepsilon$-expansion it was
shown~\cite{Halperin:1974} that no stable infrared fixed point could
exist unless the number $N$ of components of the order-parameter was
artificially extended to $N > N_c = 365$, far beyond the physically
relevant case of $N=2$. Consequently, the conclusion was that gauge
field fluctuations change the order of the phase transition to first
order \emph{irrespective} of the value of $\kappa$.

These predictions were difficult to test experimentally on
superconductors since the predicted jump across the first order
transition is very small in \emph{physical} units, even if the
\emph{effective} theory in \Eqref{GLM} has a strong first order
transition.  See e.g. Appendix A in \Citeref{Kajantie:1998hn}.  For
conventional superconductors the critical region where mean-field
behavior breaks down is extremely narrow, consequently it is very
difficult to distinguish a small finite jump from continuous
behavior.  However, there exists an isomorphism between the phase
transition in superconductors and the smectic-A to nematic transition
in liquid crystals.~\cite{Halperin:1974b} On the latter systems
experiments can be carried out in the critical
regime,~\cite{Garland:1994} and second order phase transitions are
found. This contradicts the $\epsilon$-expansion argument above, and
presumably indicates a breakdown of the expansion for this gauge-field
theory, since $\varepsilon=4-d=1$.

In \Citeref{Dasgupta:1981} it was shown, using duality arguments
and Monte Carlo simulations, that the GL model should have a second
order transition for large $\kappa$. However, what remains true is
that deep in the \typeI regime, the transition \emph{is}
first order. There should therefore be a tricritical point 
$\kappa = \kct$ where the transition changes order.

A first estimate for $\kct$ was obtained by Kleinert in
Refs.~\onlinecite{Kleinert:1982dz,Kleinert:1989book1} by developing a 
disorder theory formulation from which he calculated the value
\begin{equation*}
  \kct = \frac{3\sqrt{3}}{2\pi} \sqrt{1 - \frac{4}{9} \left(
      \frac{\pi}{3} \right)^4} \approx \frac{0.798}{\sqrt{2}}
\end{equation*}
analytically\footnote{We acknowledge Prof. H. Kleinert for pointing out
  this formula to us.}. Subsequently~\cite{Bartholomew:1983} this
picture of a tricritical point separating first and second order
transitions was given further support by Monte-Carlo simulations, and
moreover an attempt was even made to determine $\kct$, giving $\kct
\simeq 0.4/\sqrt{2}$.  However, the problem turns out to be extremely
demanding even by present day supercomputing standards, and not too
much emphasis can be put on the \emph{precise numerical value}
obtained in this early attempt. To our knowledge, this is the most
recent attempt to find a precise value for $\kct$ numerically,
although large-scale simulations have been performed much more
recently for $\kappa^2 = 0.0463$ and $\kappa^2=2$, giving first order
and continuous transitions,
respectively.~\cite{Kajantie:1998hn,Kajantie:1998vc}

The one-loop $\varepsilon$-expansion result of
Halperin~\etal~\cite{Halperin:1974} has subsequently been improved to
two-loop order,~\cite{Kolnberger:1990} drastically reducing the value
of $N_c$ to $32$, but still $N_c \gg 2$.  Eventually, an infrared
stable fixed point was found even for the physical case $N = 2$ by
combining two-loop perturbative results with Pad\'{e}-Borel
resummation techniques.~\cite{Folk:1998} {}From this latter work one can
also get an estimate of the critical $\kappa$ from
$\kappa^{\ast}=\sqrt{u^{\ast}/6f^{\ast}}\approx 0.62/\sqrt{2}$.  Since
Pad\'{e}-Borel techniques are rather uncontrolled, only simulations
can tell if such a resummation is allowed here.

{}From the above we can conclude that a tricritical $\kappa$, separating
first and second order transitions \emph{exists}, however a
\emph{precise value} remains to be determined..

We would also like to mention the distinction between \typeI and
\typeII superconductors, which is related to the response to an
external magnetic field.  When an external field is increased beyond a
critical field $H_c$ it enters a \typeI superconductor, and
superconductivity is destroyed. For \typeII superconductors the
magnetic field enters as \emph{a flux line lattice} when $H>H_{c1}$,
and superconductivity is still present in this mixed state. At the
mean-field level \typeI and \typeII superconductors are differentiated
by $\kappa = 1/\sqrt{2}$. However there is \'{a} priori no reason to
assume that this numerical value is robust against fluctuation
effects, and we will argue that the critical $\kappa$ separating first
and second order phase transitions coincides with the $\kappa$
separating \typeI and \typeII superconductors at $y_c$.

%**************************************************************************************

\section{The order of the transition}

The model in \Eqref{GLM} has a phase transition for $y=y_c$.  For
$y<y_c$ the system is in its superconducting (broken) phase while for
$y > y_c$ it is in the normal (symmetric) phase.  Note that here,
broken/symmetric does not refer to a breakdown of the local gauge
symmetry present in \Eqref{GLM}.  Elitzur's
theorem~\cite{Elitzur:1975} states that a local symmetry can never be
spontaneously broken and therefore no local order parameter (in
general any non-gauge invariant order parameter) can exist.  On the
other hand, one can explicitly break the gauge symmetry by a
gauge-fixing, thereby facilitating a meaningful definition of a local
order parameter. This should nonetheless be chosen in a formally
gauge-invariant manner to get gauge-independent results.  In our
simulations, we have chosen not to fix the gauge%
\footnote{In perturbation theory it is \emph{necessary} to fix a gauge
  to avoid infinities from the infinite number of physically
  equivalent configurations. For simulations only the explicitly
  sampled configurations contribute, and no infinities arise. Finally
  the implementation on a parallel computer is simplest without gauge
  fixing.}.%
In this case a phase transition must be found either by using
\emph{non-local}~\cite{Kajantie:1998zn,Kajantie:1998vc} order
parameters or by looking for non-analytic behavior in local
quantities,~\cite{Kajantie:1998hn} as we have done. E.g. the quantity
$\langle |\psi|^2 \rangle$ will have a jump at a first order
transition, but it will not disappear in the symmetric phase as a
proper order parameter should.  At a second order transition there
will be no jump, but the susceptibility $\chi_{|\psi|^2}$ will still
have a peak.

In principle, we could therefore decide the order by looking for a jump
in some local quantity as $\langle |\psi|^2 \rangle$, but  in finite 
systems the discontinuity will be rounded. In our case this is particularly 
problematic since the first order transitions are very weak, giving small 
jumps, even in infinite systems. At a first order transition ordered and 
disordered phases coexist and have the same free energy.  In a finite system 
there will therefore be oscillations between the different phases. Because 
of the surface energy between the two pure states the probability of finding
the system in an intermediate mixed state is lower than for either of
the pure states, and histograms of an arbitrary observable will show a
pronounced double peak structure.  This is in contrast to a second
order transition where the diverging correlation length forbids
coexistence since the whole system is correlated. The histograms then
have a single peak.  Typical histograms are shown in \Figref{fig.1}.

Thus, when these histograms have a double peak structure which becomes
more pronounced when the system size increases, the transition is
first order, otherwise not.~\cite{Lee:1990ti}

More precisely, we have the following scaling for the difference in free energy
between the mixed and pure phases for sufficiently large $L>L_{\text{scaling}}$
\begin{equation}
\Delta F(L) =
\ln{P(X,L)_{\text{max}}}-\ln{P(X,L)_{\text{min}}}
\sim L^{d-1},
\label{histscaling}
\end{equation}
where $P(X,L)$ is the probability for a given observable $X$ in a
system of size $L^d$, and $L^{d-1}$ is the cross-sectional area
between the ordered and the disordered phase.  Near the tricritical
value of $\kappa$ such scaling is difficult to achieve since we are
interested in the limit of vanishingly weak first order transitions.
Consequently, a very large $L$ is required in order to observe proper
scaling. Only for quite strong first order transitions have we been
able to observe proper scaling as predicted by \Eqref{histscaling},
however we have generally taken a monotonous increase in $\Delta F(L)$
with system size as a signature of a first order phase transition. For
the weakest first order transitions $\Delta F(L)$ will typically
decrease for small $L$ and then start to increase. It is therefore
important to observe monotonic behavior through several system sizes
before a conclusion can be drawn from the histograms\footnote{The
  proportionality factor between $\Delta F$ and the cross-sectional
  area $L^{d-1}$ is the surface tension $\sigma$, which vanishes at
  $\xct$, but due to the difficulty in getting proper $L^{d-1}$
  scaling of $\Delta F$, we have not considered $\sigma$.}.

%*****************************************************************************

\section{Phase diagram}
We are searching for the point in the $(x,y)$ plane where a first
order and a second order line meet, i.e. according to the rather
loose definition\footnote{Indeed, it has been customary to describe as
  ``tricritical'' any point at which a continuous transition becomes
  discontinuous, irrespective of the number of phases which coexist
  along the first-order line or of the number of lines or surfaces of
  ordinary critical points which, in a suitably enlarged parameter
  space, may be found to meet here.} of Lawrie and Sarbach~\cite{Lawrie:1984}
we are looking for a \emph{tricritical point}. At a tricritical point
\emph{two} coupling constants must be fine-tuned to nontrivial values, and 
consequently a tricritical theory can be described with the mean-field
free energy
\begin{equation}
f \approx |\nabla\psi|^2+c_1(y-\yct)|\psi|^2+c_2(x-\xct)|\psi|^4+c_3|\psi|^6.
\end{equation}
Right at the tricritical point the coefficients in front of both
$\left|\psi\right|^2$ and $\left| \psi \right|^4$ vanish
\emph{simultaneously}. The upper critical dimension for this model is
$d^{\ast}=3$ and mean-field theory should be valid (up to logarithmic
corrections). When approaching the tricritical point from the first
order side, mean-field theory predicts that the jump $\Delta \left| \psi
\right|^2$ will vanish as
\begin{equation}
 \Delta \left| \psi \right|^2 \sim \left( \xct - x \right).
\label{meanfield:scaling}
\end{equation}

We will make use of the above scaling in section \ref{section:results} to
estimate $\xct$. For further information about tricritical points, we
refer to an extensive review by Lawrie and Sarbach.~\cite{Lawrie:1984}

\begin{figure}
       \includegraphics[width=3.1in]{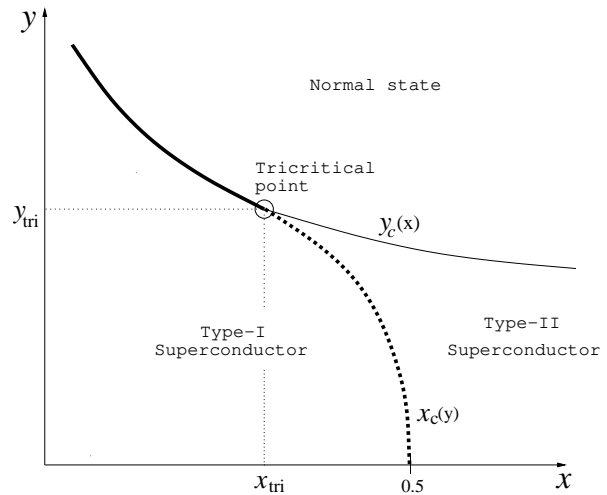}
        \caption{A conjectured phase diagram in the $(x,y)$ plane
          \emph{in the vicinity of the tricritical point}.  The thick
          solid line is a line of first order transitions separating
          \typeI superconductivity and the normal (metallic) state,
          the thin solid line is a second order line separating
          \typeII superconductivity from the normal (metallic) state.
          The dashed line separates \typeI and \typeII
          superconductivity.  The dotted horizontal and vertical lines
          indicate the coordinates of the tricritical point
          $(\xct,\yct) \simeq (0.30,0.03)$.  \label{fig:phase} }
\end{figure} 

In \Figref{fig:phase} we have \emph{assumed} that the tricritical
point separating first order and second order phase transitions
coincides with the point separating \typeI and \typeII
superconductivity. In principle the line of second order transitions
could extend into the \typeI region, with an intermediate state of
\typeI superconductivity with a second order phase transition to the
normal state. This would be the case if the mean field value
$\kappa_{\text{I}/\text{II}} = 1/\sqrt{2}$ was \emph{not} renormalized
by fluctuations. We have not focused on the aspect of \typeIII
superconductivity in our simulations, we will however argue that the
overall structure of the phase diagram shown in \Figref{fig:phase} is 
correct in the vicinity of the tricritical point.

The microscopic difference between \typeI and \typeII superconductors
lies in the sign of the effective vortex-vortex interaction. 
In $d=3$ there exists a {\it dual} formulation of the GL-model which
is given by a complex scalar matter-field $\phi$ coupled minimally to
a massive gauge-field. This gauge-field can thus safely be integrated
out to yield an effective \emph{local} $|\phi|^4$-theory, where the
coefficient of the $|\phi|^4$-term gives the effective vortex-vortex
interaction.  A positive such term signals vortex-repulsion, i.e.
type-II behavior, while a negative term signals type-I behavior.
\emph{This vortex-vortex interaction term is proportional to
  $\kappa-\kct$} where $\kct$ is indeed to be identified with our
tricritical value of $\kappa$.~\cite{Kleinert:1982dz} Using the dual
formulation of the GL theory, it then becomes clear that $\kct$
is at the same time the value that separates first order and second
order behavior, \emph{and} the value that separates attractive from
repulsive effective vortex-vortex interactions, i.e. type-I from
type-II behavior.

An independent argument for why the transition between the normal
state and \typeI superconductivity \emph{must be first order}, is
based on the geometrical properties of a vortex tangle: In a recent
paper~\cite{Hove:2000hausd} we have calculated the fractal dimension of
vortex loops, and found the scaling relation $ \beta = \nu\left(d -
  D_{\text{H}}\right)/2$, where $\beta$ is the order parameter
exponent, $\nu$ is the correlation length exponent and $D_{\text{H}}$ is the
fractal dimension of the loops. If we formally extend this relation to
the first order regime, i.e. let $\beta \to 0^+$, we find that the
fractal dimension of the vortex loops $D_{\text{H}} \to d$, i.e. the
vortices collapse on themselves (filling space completely), rendering
the transition discontinuous.  This collapse is what we would expect
from vortices interacting attractively (i.e.  \typeI), and by turning
the argument above around we conclude that \typeI superconductors must
have a first order transition to the normal state.

We emphasize that the detailed \emph{shape} of the line $x_c(y)$
remains to be determined. We have presented arguments above that it
ends in the tricritical point $(\xct,\yct)$. Moreover, deep in
the broken regime, mean-field theory should apply. Consequently, we expect
that the line $x_c(y)$ converges towards the mean field value
$x_{\text{I}/\text{II}} = 1/2$ in the $y \to -\infty$ limit. 

%************************************************************************************
\section{Lattice Model}

{}To perform simulations on the model in \Eqref{GLM} we define it
on a numerical lattice of size $N \times N \times N$ with
lattice constant $a$. The \emph{physical} volume is then
$V=L^3=\left(Na\right)^3$. By introducing a lattice field given by
\begin{equation}
\label{eq:fieldscaling}
        |\phi_{\text{(cont)}}|^2 = \beta_H |\psi_{\text{(latt)}}|^2/2a,
\end{equation}
where $\beta_H$ so far is an arbitrary constant, \Eqref{GLM} takes the form 
\begin{align}
 Z &= \int \mathcal{D} A_i \mathcal{D}\psi \exp(-S(A_i,\psi)) \nonumber \\
 S &= \beta_G \sum_{\vec{x},i<j} \frac{1}{2}F_{ij}^2 
     -\beta_H \sum_{\vec{x},i} \mathsf{Re} \biggl(\psi^{\ast}(\vec{x}) U_i(\vec{x}) 
                                                  \psi(\vec{x} + \hat{\imath}) \biggr) \nonumber \\
   &\quad +\frac{\beta_H}{2}\left[6+\frac{y}{\beta_G^2}\right]\sum_{\vec{x}} |\psi|^2
     +\beta_R \sum_{\vec{x}} |\psi|^4   
\end{align}
where we have defined $\alpha_i(\vec{x}) = a q A_i(\vec{x})$,
$U_i(\vec{x}) = e^{i \alpha_i(\vec{x})}$, $\beta_G = 1/\ladeta$,
$F_{ij}=\alpha_i(\vec{x})+\alpha_j(\vec{x}+\hat{\imath})
-\alpha_j(\vec{x})-\alpha_j(\vec{x}+\hat{\jmath})$, and $\beta_R = x
\beta_H^2/4 \beta_G$.  $F_{ij}$ is essentially a lattice curl of the
fluctuating gauge-field, and $\ladeta = a\charge^2$ is a
dimensionless lattice constant. To obtain correct continuum limit results,
we will ultimately be interested in the limit $\ladeta \to 0$. It is
furthermore possible to select a value of $\beta_H$ such that the action can be
written on the form
\begin{align}
S=\beta_G \sum_{\vec{x},i<j} \frac{1}{2}F_{ij}^2 
&- \beta_H  \sum_{\vec{x},i} \mathsf{Re} \biggl(\psi^{\ast}(\vec{x})U_i(\vec{x}) 
\psi(\vec{x} + \hat{\imath}) \biggr) \nonumber \\
&+ \sum_{\vec{x}} |\psi|^2 + \beta_R \sum_{\vec{x}} [|\psi|^2-1]^2.
\label{GLML}
\end{align}
This is achieved provided $\beta_H $ satisfies the relation $(\beta_H/2)[6+y/\beta_G^2] + 
2 \beta_R=1$. 

The amplitude and gauge-invariant phase difference $\Delta =
\mathsf{arg} \bigl(\psi^{\ast}(\vec{x})U_i(\vec{x}) \psi(\vec{x} +
\hat{\imath}) \bigr)$ are coupled through the second term in
\Eqref{GLML}. The ordered state is characterized by $\cos \Delta
\lesssim 1$ and $\left|\psi\right|$ close to the minimum in the
potential energy, whereas in the disordered state $\cos \Delta \approx
0$. In the disordered state the amplitude behavior is determined by
$x$; for small $x$ the coupling to $\Delta$ dominates and
$\left| \psi \right|$ deviates significantly from the minimum in the
potential, whereas for large $x$ amplitude fluctuations are
suppressed.

Given the fact that the theory in \Eqref{GLM} is a continuum theory, one 
has to perform an ultraviolet (short-distance) renormalization, and thus
$m^2=m^2(\charge^2)$ has to be interpreted as a renormalized mass
parameter at a given scale $\charge^2$ within a given renormalization
scheme, e.g. the minimal subtraction ($\overline{MS}$) scheme.  Since
this continuum theory should represent the $a\rightarrow 0$ limit of the
lattice theory in \Eqref{GLML}, the parameter $y$ must be varied when
$a$ is being varied. In our case the leading terms in $a$ can be obtained 
by requiring that some physical correlator calculated in both lattice and
continuum perturbation theory should coincide. Thus we have to make the
substitution~\cite{Laine:1995ag,Laine:1998dy}
\begin{align}
y \to y &- \frac{3.1759115(1+2x)}{2\pi}\beta_G \notag \\
        &- \frac{\left(-4+8x-8x^2\right)\left(\ln{(6\beta_G)}+0.09\right)-1.1+4.6x}{16\pi^2} \notag \\
        &+ \mathcal{O}(1/\beta_G)
\label{RG:y}
\end{align}
In addition, the continuum and lattice condensate matter fields are related by
\begin{equation}
\begin{split}
\frac{\langle\phi^{\ast}\phi\rangle_{\text{cont}}}{q^2} 
&= \frac{\beta_H\beta_G}{2} \langle\psi^* \psi\rangle_{\text{latt}} \\
&-\frac{3.175911 \beta_G}{4 \pi} - \frac{\log(6 \beta_G)+0.668}{8\pi^2}
+\mathcal{O}(1/\beta_G).
\end{split}
\label{RG:condensate}
\end{equation}
In \Eqref{RG:condensate} the first term comes from \Eqref{eq:fieldscaling}, 
while the second and third terms are linear and logarithmic divergences due 
to renormalization.

Note that the complicated counterterms in \Eqref{RG:y} merely affect the \emph{value} 
of $y_c$ separating the normal from the superconducting state for a given $x$, not 
the overall \emph{structure} of the phase-diagram. The divergences in 
\Eqref{RG:condensate} in the continuum limit are constants that cancel when the 
jump in $\langle \phi^{\ast}\phi\rangle$ across a first order phase transition
is calculated.

%******************************************************************************
\section{Details of simulations}

In order to use \Eqref{GLML} to study the continuum theory of
\Eqref{GLM}, it is necessary to carefully take two limits separately.
First, the infinite volume limit $L \to \infty$ is taken, thereafter the
continuum limit $a \to 0$.  For reliable results one should have $a
\ll \xi \ll L$, where $\xi$ is a typical correlation length for the
problem. In statistical physics, the continuum limit is usually not
considered, either because the models are inherently \emph{lattice
models}, or the models are studied around a second order critical
point where there exists at least one diverging length scale. Under such
circumstances the short length-scale properties, like the lattice constant, 
are rendered irrelevant when studying universal properties. On the other hand, 
if one wants to study non-universal properties (such as critical coupling constants) 
or first order transitions without a diverging length scale, details of
the system even on the shortest length scales have to be correctly
taken into account in order to give reliable results.

The Monte-Carlo simulations are performed on \Eqref{GLML}, updating phases,
amplitudes%
\footnote{%
  Note that in the London limit, with spatially constant amplitude,
  one cannot access the \typeI regime. The 3DXY model coupled to a
  gauge-field is the dual of the 3DXY-model with no gauge-field
  fluctuations.  The latter has a critical point corresponding to the
  3DXY universality class, while the former recently has been shown
  explicitly to have a stable infrared charged fixed point. See J.
  Hove and A. Sudb{\o}, Phys. Rev. Lett., {\bf 84}, 3426 (2000).}, and
gauge-fields.  We have used periodic boundary conditions and
non-compact gauge-fields without any gauge fixing.  To reduce
autocorrelation times we have added global updating of the amplitude
and overrelaxation of the scalar
field~\cite{Dimopoulos:1999aa,Kajantie:1996kf} such that one sweep
consists of: (1) conventional local Metropolis updates for phase,
amplitude and gauge field, (2) global radial update by multiplying the
amplitude uniformly with a common factor (acceptance according to
Metropolis dynamics) and (3) 2-3 overrelaxation ``sweeps'' updating
both the amplitude and the phase of the scalar field.  The acceptance
ratio in the Metropolis steps is kept between 60-70\% as long as
possible by adaptively adjusting the maximum allowed changes in the
fields.  For further details of the technical aspects of the
simulations, see Refs.~\onlinecite{Dimopoulos:1999aa,Kajantie:1996kf}.

We have performed simulations for the parameters in Table~\ref{tab.sim}.
\begingroup
\squeezetable
\begin{table}
\caption{
The lattice sizes $N^3$ used for each $(\ladeta,x)$-pair. For each 
lattice size typically between three and eight $y$-values were used.
The symbols are defined by:\newline
\ns$\,$   Not simulated\newline
\sok$\,$  Simulated \newline
\ssim$\,$ Simulated and results shown in \Figref{fig.1}.} 

\label{tab.sim}
\begin{ruledtabular}
\begin{tabular}{c|c|c|c|c|c|c|c|c|c|c|c|}
$\ladeta$ & $x$ & \multicolumn{10}{c}{$N$}\vline \\
\cline{3-12}
     &       &~8&12&16&20&24&32&40&48&64&96 \\
\colrule
5.0\footnote{In Ref.~\onlinecite{Bartholomew:1983}
the lattice spacing corresponds to $\ladeta=5.0$.
The system sizes used were $9^3$ and $15^3$.}
     & 0.10                   & \sok & \ns  & \ns   & \ns   & \ns   & \ns   & \ns    & \ns   & \ns   & \ns  \\
     & 0.15                   & \sok & \sok & \sok  & \sok  & \sok  & \sok  & \ns    & \ns   & \ns   & \ns  \\
     & 0.16                   & \sok & \sok & \ssim & \ssim & \ssim & \ssim & \ns    & \ns   & \ns   & \ns  \\
     & 0.17, 0.18, 0.19       & \sok & \sok & \ssim & \ns   & \ssim & \ssim & \ssim  & \ns   & \ns   & \ns  \\
\colrule                                         
2.0  & 0.10                   & \sok & \sok & \ns   & \ns   & \ns   & \ns   & \ns    & \ns   & \ns   & \ns  \\
     & 0.15                   & \sok & \sok & \sok  & \ns   & \ns   & \ns   & \ns    & \ns   & \ns   & \ns  \\
     & 0.20                   & \sok & \sok & \sok  & \ns   & \sok  & \sok  & \sok   & \ns   & \ns   & \ns  \\
     & 0.22                   & \ns  & \ns  & \ssim & \ns   & \ssim & \ssim & \ssim  & \ns   & \ns   & \ns  \\
     & 0.23, 0.24, 0.25       & \ns  & \ns  & \ssim & \ns   & \ssim & \ssim & \ssim  & \ns   & \ssim & \ns  \\
\colrule                   
1.0  & 0.08                   & \sok & \sok & \sok  & \ns   & \ns   & \ns   & \ns    & \ns   & \ns   & \ns  \\
     & 0.10                   & \sok & \sok & \sok  & \sok  & \ns   & \ns   & \ns    & \ns   & \ns   & \ns  \\
     & 0.12, 0.13, 0.14       & \sok & \sok & \sok  & \ns   & \sok  & \sok  & \ns    & \ns   & \ns   & \ns  \\
     & 0.15, 0.16, 0.17       & \ns  & \sok & \sok  & \sok  & \sok  & \sok  & \sok   & \ns   & \ns   & \ns  \\
     & 0.18, 0.20             & \sok & \sok & \sok  & \ns   & \sok  & \sok  & \sok   & \sok  & \ns   & \ns  \\
     & 0.22                   & \sok & \sok & \sok  & \ns   & \sok  & \sok  & \sok   & \sok  & \sok  & \ns  \\
     & 0.24, 0.25, 0.26, 0.27 & \sok & \sok & \ssim & \ns   & \ssim & \ssim & \ssim  & \ssim & \ssim & \ns  \\
     & 0.30                   & \sok & \sok & \sok  & \ns   & \sok  & \sok  & \sok   & \sok  & \ns   & \ns  \\
     & 0.50                   & \ns  & \ns  & \sok  & \sok  & \sok  & \sok  & \sok   & \ns   & \ns   & \ns  \\
\colrule                 
0.5  & 0.16                   & \ns  & \ns  & \ns   & \ns   & \ssim & \ssim & \ssim  & \ssim & \ns   & \ns  \\
     & 0.20, 0.24             & \ns  & \ns  & \ns   & \ns   & \ssim & \ssim & \ssim  & \ssim & \ssim & \ns  \\
     & 0.26, 0.28             & \ns  & \ns  & \ns   & \ns   & \ssim & \ssim & \ssim  & \ssim & \ssim & \ssim  \\
     & 0.30                   & \ns  & \ns  & \ns   & \ns   & \ssim & \ssim & \ssim  & \ssim & \ssim & \ns  \\
\end{tabular}

\end{ruledtabular}
\end{table}
\endgroup 
The simulations have been done in a hierarchical manner: For a given
$x$ we have first kept $\ladeta$ and $N$ fixed, and simulated on
typically three to eight $y$ values. These runs have been combined
with Ferrenberg-Swendsen~\cite{Ferrenberg:1988yz,Ferrenberg:1989ui}
reweighting techniques, and a (pseudo)critical $y$ has been located by
requiring that the reweighted histograms have two equally high%
\footnote{%
  We have used equal \emph{height} histograms instead of equal
  \emph{weight}. The reason for this is that, in particular for small
  systems, the histograms are quite asymmetric. Then the equal weight
  histograms are not well defined for weak first order transitions.
  Both methods should give the same results in the infinite volume
  limit, but the convergence rate may be different.} peaks. Then the
  system size has been increased to access the \emph{infinite volume
  limit}, and finally we have varied $\ladeta$ to determine the
\emph{continuum limit}. %
At the transition the number of sweeps was chosen so that the system
oscillated back and forth between the ordered and disordered state
about ten times. Depending on system size and $x$-value (i.e. the
\emph{strength} of the first order transition) this resulted in about
$10^5$ to $10^6$ sweeps.  All computations were performed on an SGI
Origin 3800 at the Norwegian High Performance Computing Center, using
up to 32 nodes in parallel for the largest systems. A total of about
$5\cdot 10^4$ CPU hours were used, corresponding to $ \simeq 1.5 \cdot
10^{17}$ floating point operations.

\section{Results}
\label{section:results}
{}To find $\xct=\kct{}^2$ our strategy has been to start at $x\ll
\xct$ where the transition is clearly first order, and then slowly
increase $x$ into the problematic tricritical area where $x\lesssim
\xct$. During the simulations we have sampled the
lattice amplitude
\begin{equation}
\RR = \frac{1}{N^3}\sum_{\vec{x}} |\psi(\vec{x})|^2
\label{simulation:r2def}
\end{equation}
and histograms of this quantity constitute the raw data for most of the
subsequent analysis%
\footnote{In addition to $\RR$ we have also studied other quantities, in
particular
\[
\overline{L} = \frac{1}{3N^3} \sum_{\bvec{x},i} \cos \left(\mathsf{arg}\left[ \psi^{\ast}(\vec{x}) U_i(\vec{x})
\psi(\vec{x}+\hat{\imath}) \right]\right),
\]
which varies between zero in the symmetric state and one in the broken
state, quite similar to the more familiar 3DXY quantity $\langle \cos
\left( \theta(\bvec{x}) - \theta(\bvec{x} + \hat{\imath})\right)
\rangle$. However, the general picture is that the different
observables give essentially the same information, and we have
therefore focused mainly on \Eqref{simulation:r2def}, which has a 
well defined continuum limit.}. The connection between continuum and
lattice condensates is given by \Eqref{RG:condensate}.

\begin{figure*}
        \includegraphics[angle=270,width=3.0in]{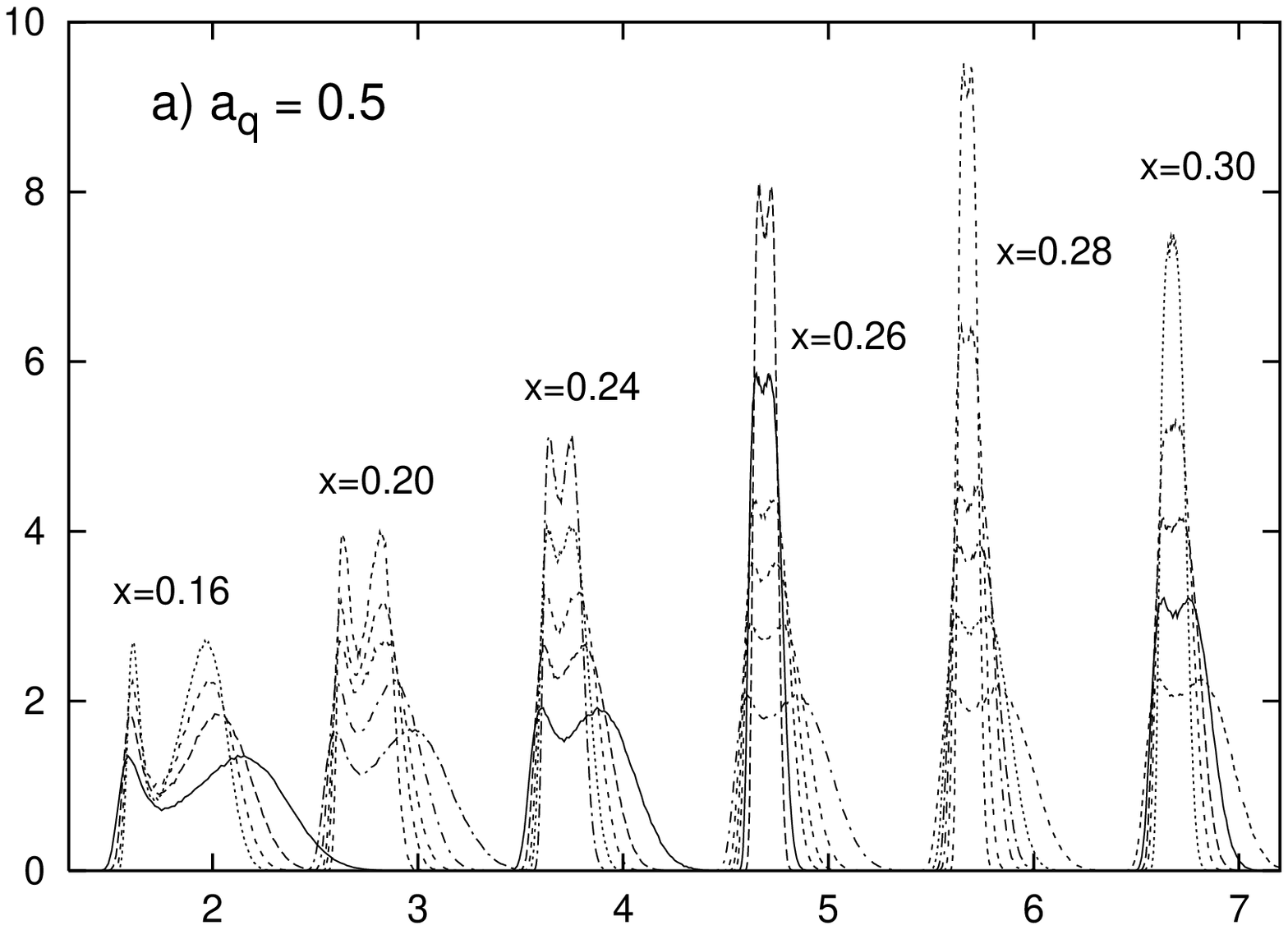}
        \includegraphics[angle=270,width=3.0in]{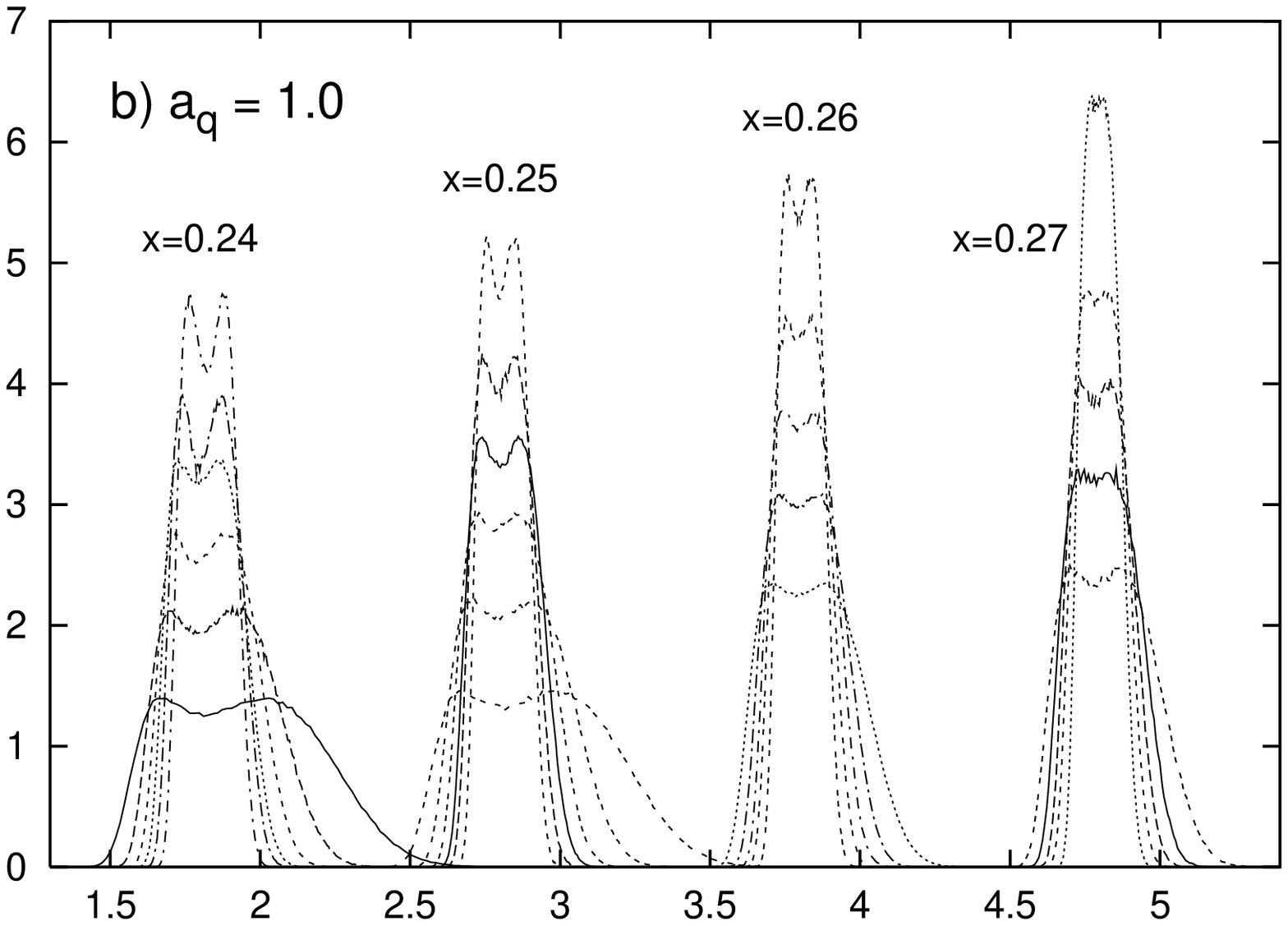}
        \includegraphics[angle=270,width=3.0in]{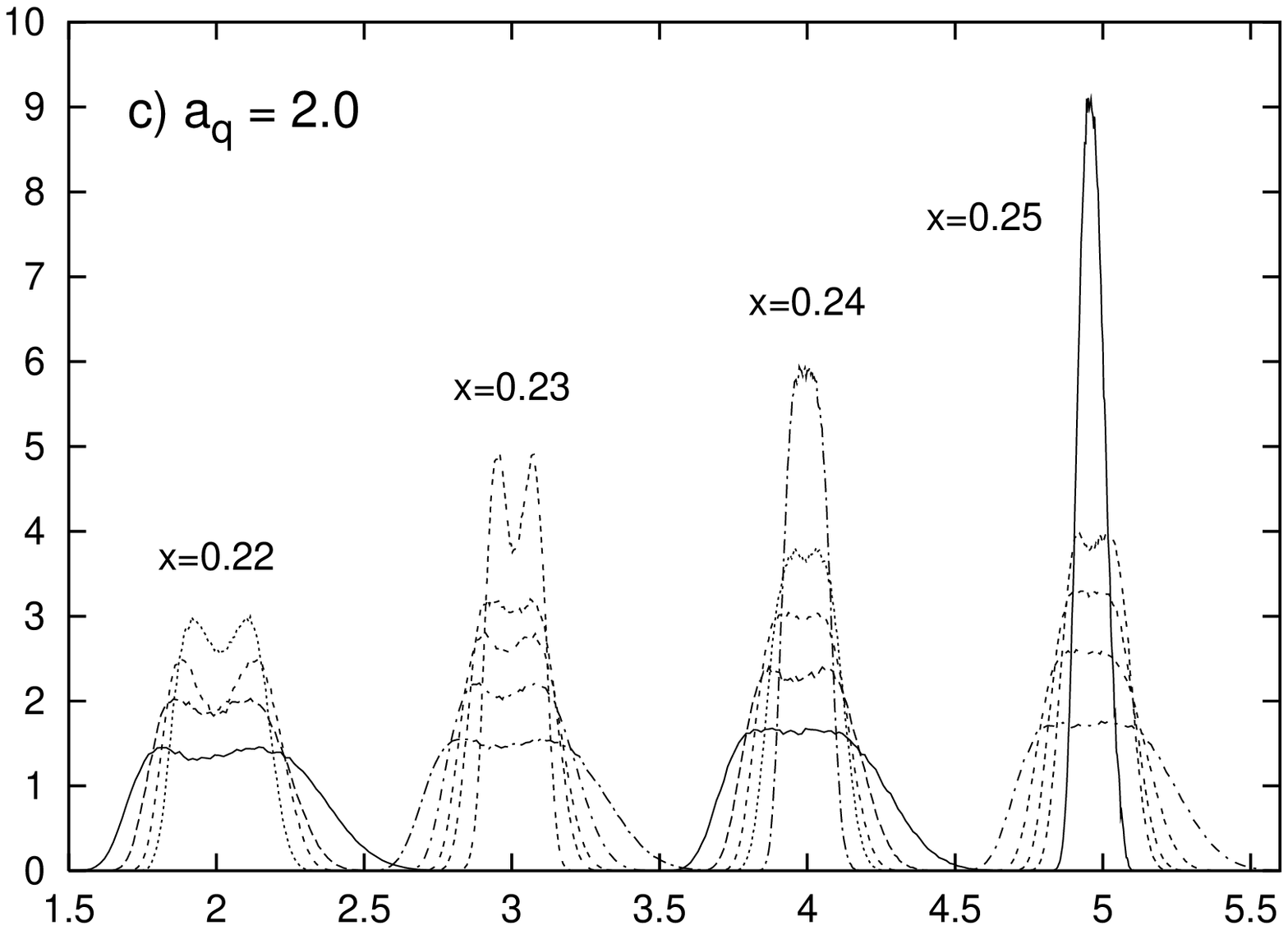}
        \includegraphics[angle=270,width=3.0in]{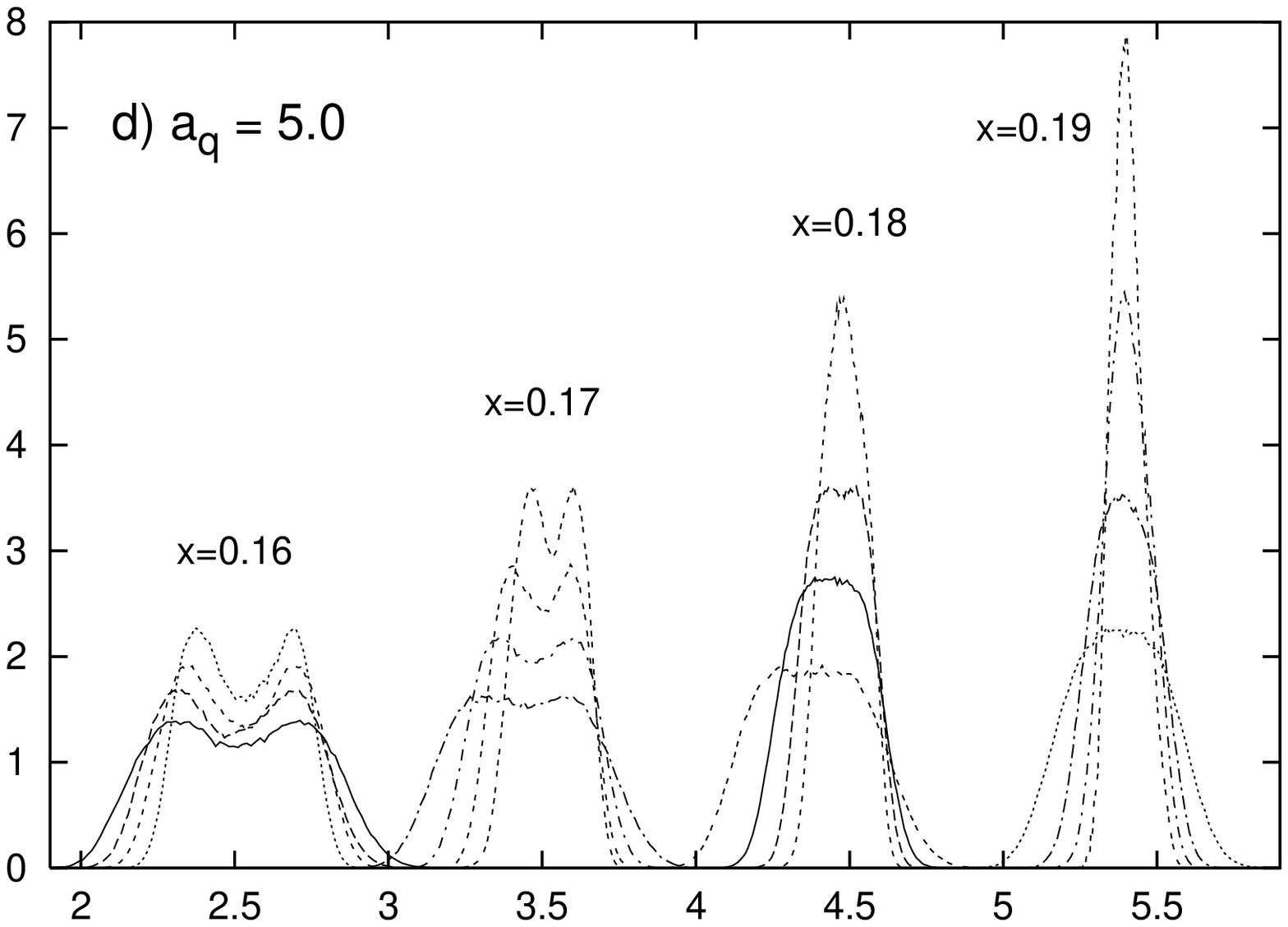}
        \caption{Normalized histograms $P(\RR)$ as a function of $\RR$
          for a) $\ladeta=0.5$, b) $\ladeta=1.0$, c) $\ladeta=2.0$, d)
          $\ladeta=5.0$.  For each lattice spacing the histograms for
          the smallest $x$ are correctly placed horizontally. For
          larger $x$ they are offset horizontally in steps of $1$ for
          clarity.  For system sizes see Table~\ref{tab.sim}.
          \label{fig.1}
        }
\end{figure*}

Histograms reweighted to the critical $y$-value are shown in
\Figref{fig.1}.
We have used two different methods to find $\xct(\ladeta)$ from the
histograms, and finally at the end of this section we have
extrapolated these values to $\ladeta=0$ to find the continuum limit.

\subsection{Extrapolation of $\deltaR$ to zero}
\begin{figure*}
        \includegraphics[angle=270,width=6in]{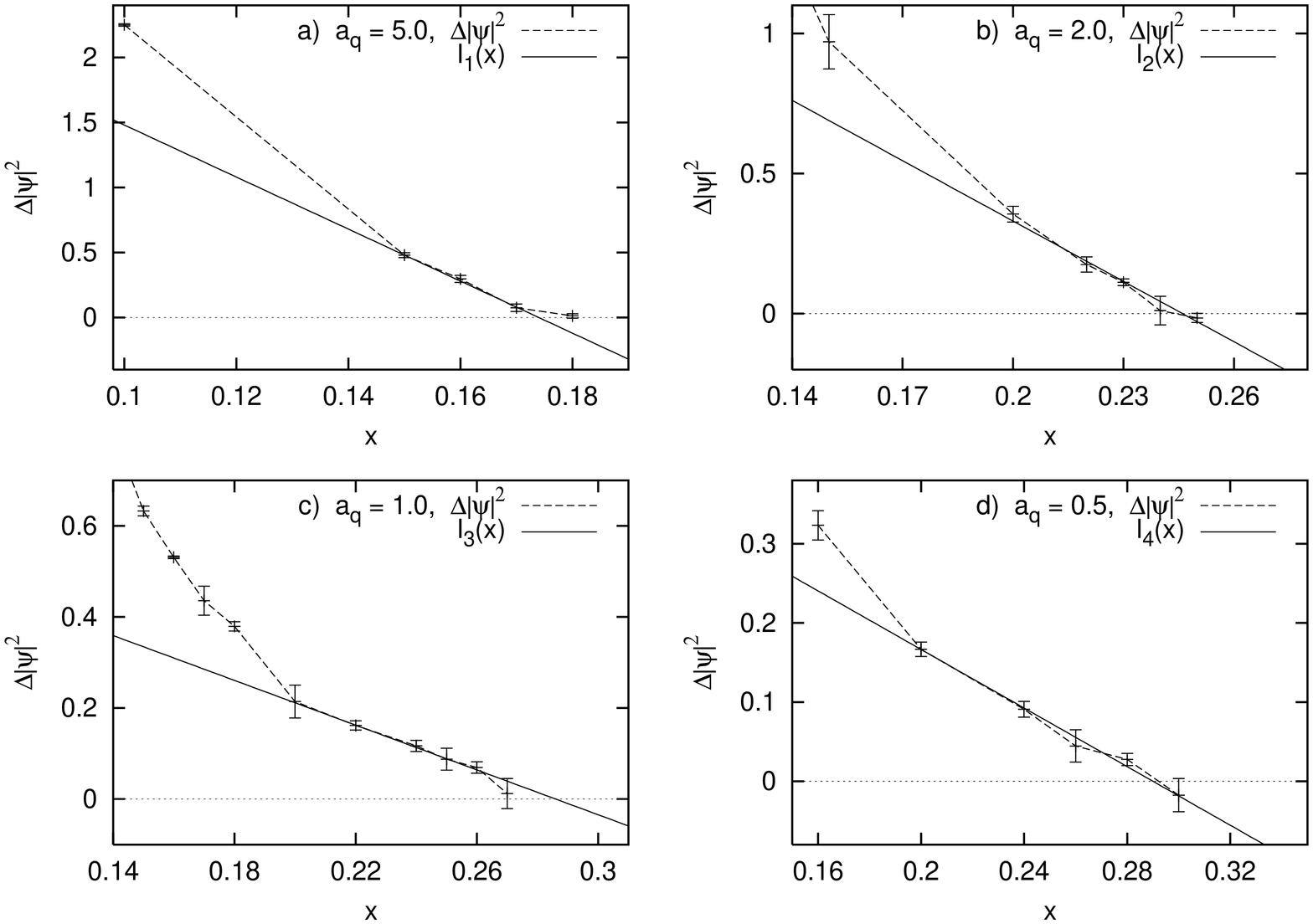}
        \caption{$\lim_{N\to\infty}\deltaR$ as a function of
                 $x=\kappa^2$ for the lattice constants
                 a) $\ladeta = 5.0 $,
                 b) $\ladeta = 2.0 $,
                 c) $\ladeta = 1.0 $,
                 d) $\ladeta = 0.5 $.
                 The line $l_i(x), i=1,\ldots,4$ is a fit to \Eqref{meanfield:scaling}
                 where $x_c$ is given in Table~\ref{tab.xc}.
                \label{fig.2}
                }
\end{figure*}

The distance between the peaks of a histogram gives $\deltaR (N)$, and by computing this for several different system
sizes one can compute the infinite volume limit
$\lim_{N\to\infty}\deltaR$ of the discontinuity at the
transition. Then one can (in principle) extrapolate to larger $x$ and
find the value $\xct$ where the discontinuity disappears. Results for
$\lim_{N\to\infty}\deltaR$ as a function of $x$ are
shown in \Figref{fig.2}.

For small $x$ the curves in \Figref{fig.2} show a distinct positive
curvature, but when approaching $\xct$ we find that $\deltaR$ vanishes as $\propto (\xct-x)$, in accordance with
mean-field theory, \Eqref{meanfield:scaling}. Also in the original
attempt to locate $\xct$ with Monte Carlo
simulations~\cite{Bartholomew:1983} this extrapolation was done,
however the extrapolation was done starting from quite small $x$
values, and the resulting $\xct$ was much smaller than the one we
calculate.

The extrapolated results for $\xct$ are shown in Table~\ref{tab.xc}.
\begin{table}
\caption{$\xct$ found from extrapolation of $\lim_{N\to\infty}\deltaR$ to zero
         and finite size scaling of $\Delta F(N)$.}
\label{tab.xc}
\begin{ruledtabular}
\begin{tabular}{ccc}
$\ladeta$ & $\xct$ (from $\deltaR$)  & $\xct$ (from $\Delta F(N)$) \\
\colrule
5.0  & 0.174$\pm$0.002  & 0.175$\pm$0.005  \\
2.0  & 0.246$\pm$0.002  & 0.235$\pm$0.005  \\
1.0  & 0.286$\pm$0.010  & 0.260$\pm$0.010  \\
0.5  & 0.294$\pm$0.005  & 0.280$\pm$0.020  \\
\end{tabular}
\end{ruledtabular}
\end{table}
The values found should provide a reasonable upper limit for $\xct(\ladeta)$.

\subsection{Finite size scaling of $\Delta F(N)$}

\begin{figure*}
        \includegraphics[angle=270,width=5.0in]{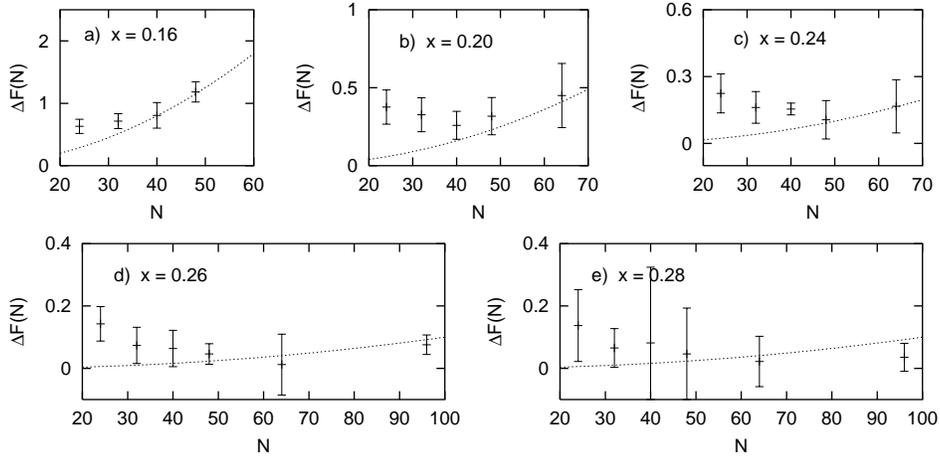}
        \caption{$\Delta F(N)=\ln{P_{\text{max}}(N)}-\ln{P_{\text{min}}}(N)$
                 for $\ladeta=0.5$. The line is $\propto N^2$ which is the 
                 scaling in \Eqref{histscaling} (for $d=3$).
                \label{fig.3}
                }
\end{figure*}

It is also possible to study the \emph{height} of the peaks in the
histograms [$P(\RR)_{\text{max}}$] relative to the minimum
between them [$P(\RR)_{\text{min}}$].  This constitutes the best method 
of determining whether a transition is first order or not, but one
cannot extrapolate to find $\xct$.  In \Figref{fig.3} we show some
typical results for $\Delta
F(N)=\ln{P_{\text{max}}}-\ln{P_{\text{min}}}$ as function of system
size $N$ for $\ladeta=0.5$.

For $x=0.16$ we clearly see the scaling $\Delta F(N) \propto N^2$
for $N\gtrsim 40$. This is expected since the histograms in
\Figref{fig.1} show a very pronounced double peak structure.
For larger $x$ this becomes less clear. Our estimates of $\xct$ for
the different lattice constants are given in Table~\ref{tab.xc}.
The results are consistently somewhat below those found
with method A and give a reasonable lower limit for $\xct$.

\subsection{Other methods}
Finite-size scaling of the maximum in susceptibilities of the 
quantities $\RR$ and $\overline{L}$ gives results
that are consistent with the above conclusions.
\begin{equation}
\chi_{S}=N^d\left( \left\langle S^2 \right\rangle-\left\langle S \right\rangle^2 \right) 
        \sim N^{\sigma},\quad S\in\{\RR,\overline{L}\}
\end{equation}
where $\sigma=d (<d)$ for first(second) order transitions.
However, these results are more ambiguous than those from
the histograms, and we have therefore chosen to work mainly
with the histograms.
 
\subsection{Final result for $\kct$}
\label{result:final}
It is clear that it becomes increasingly difficult to obtain good estimates of
$\kct(\ladeta)$ when the lattice constant is reduced. This is easy to understand
since the \emph{physical} volume $(N\ladeta)^3$ will be drastically reduced for the same 
lattice size in lattice units. The size of $N$ necessary to access the scaling regime 
is (approximately) inversely proportional to the lattice constant $\ladeta$. 

\begin{figure}
        \includegraphics[angle=270,width=3.1in]{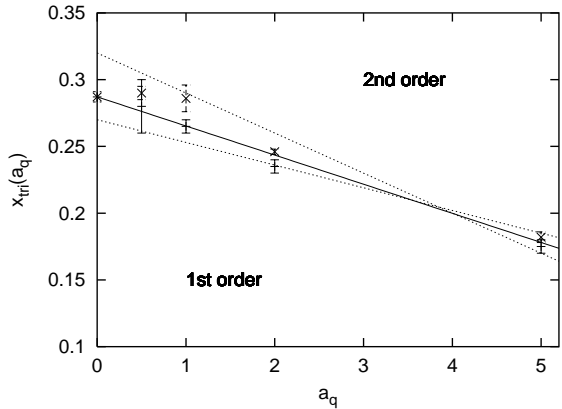}
        \caption{Plot of $\xct(\ladeta)$ using the results from extrapolation of
          $\deltaR$ to zero and from finite size scaling of $\Delta
          F(N)$ given in Table~\ref{tab.xc}. The solid line is a
          linear fit giving $\lim_{\ladeta \to 0} \xct(\ladeta) =
          0.287\pm 0.004$. The dotted lines indicate ``worst case
          scenarios'' giving $\lim_{\ladeta \to 0} \xct(\ladeta) =
          0.295\pm 0.025$.
          \label{fig.4}
        }
\end{figure}

In \Figref{fig.4}, we show $\xct(\ladeta)$ found from extrapolation of
$\deltaR$ to zero and from finite size scaling of
$\Delta F(N)$ as given in Table~\ref{tab.xc}.  
A linear fit to the data gives 
$\lim_{\ladeta \to 0} \xct(\ladeta) = 0.287\pm 0.004$
with a confidence level of 25\%. This is probably an underestimate
of the error, since we have no particular reason to assume
a linear behavior. Since the errors in $\xct(\ladeta)$
increases considerably when we reduce $a_q$ one cannot
rule out other behaviors, as quadratic. 
{}From the ``worst case scenario'' shown
by the dotted lines in the figure we get 
$\lim_{\ladeta \to 0} \xct(\ladeta) = 0.295\pm 0.025$.
This in all likelihood gives a more realistic estimate of the error,
and we therefore give our final estimate of $\kct$ as
$\lim_{\ladeta\to 0} \kct(\ladeta) = (0.76\pm 0.04)/\sqrt{2}$.

%-------------------------------------------------------------------------
\section{Conclusion}

In summary, we have presented results from large scale Monte Carlo
simulations showing that the critical value of the Ginzburg-Landau
parameter that separates first order from second order behavior at the
superconductor-normal metal transition point, is $\kct = (0.76\pm
0.04)/\sqrt{2}$. This is in remarkable agreement with the
first estimate  of $\kct$ obtained by Kleinert~\cite{Kleinert:1982dz} 
using a mean-field theory on the dual of the Ginzburg-Landau model, 
but differs almost by a factor of two from the subsequent early
simulation results of Bartholomew.~\cite{Bartholomew:1983}

The reason for the remarkable agreement with our result and those of
\Citeref{Kleinert:1982dz}, is that for small to intermediate
values of $\kappa$, the original problem is in the strong coupling
regime and is mapped onto a weak-coupling problem in the dual
formulation. The dual model is then expected to yield rather precise
results at the mean-field level.~\cite{Kleinert:1982dz} The dual
description of the Ginzburg Landau model has recently met with
considerable success in predicting the phase-structure of extreme
type-II superconductors, even in magnetic fields.~%
\cite{Tesanovic:1999,Nguyen:1999epl,Nguyen:1999prb,Hove:2000etaA} We
interpret the good agreement between our results and those of
\Citeref{Kleinert:1982dz} as further support to the dual
description of the Ginzburg-Landau model, now also in the intermediate-$\kappa$ 
region.

We have also argued that this $\kct$ coincides with the $\kappa$
separating \typeI and \typeII superconductivity. In the
superconducting regime for $\kappa \in (\kct,1/\sqrt{2})$ we thus
predict the possibility of going from \typeI to \typeII
superconductivity by increasing the temperature. This could in principle 
be possible to observe by studying the vortex structure of a superconductor 
with such intermediate values of $\kappa$ by small-angle neutron scattering, 
when lowering the temperature 
through the line $x_c(y)$. However, more work is needed to elucidate the 
properties of the line $x_c(y)$ in \Figref{fig:phase}. 

\begin{acknowledgments}
We especially thank Dr.~K.~Rummukainen for generously providing us 
with the software for the data analysis, and for numerous helpful discussions 
during a visit by two of us (S.M. and J.H.) to Nordita, and later.
We also thank Prof.~H.~Kleinert,  Dr.~F.~Nogueira, and Prof.~Z.~Te{\v s}anovi\'{c} 
for useful discussions. We also thank all of the above for critical readings of the manuscript.
A. S. thanks H. Kleinert and the Freie Universit{\"a}t Berlin for their hospitality 
while this work was  being completed. This work was supported by the Norwegian Research 
Council via the High Performance Computing Program and  Grant No.~124106/410 (S.M. and A.S.), 
and by NTNU through a fellowship (J.H.).

\end{acknowledgments}

\end{document}